%% file: draft.tex
\documentclass[oldversion]{aa}  

\makeatletter
\let\linenumbers\relax
\let\runningpagewiselinenumbers\relax

\let\@LN@col\@gobble

\makeatother

\newcommand{\lens}{MACS\,0416}
\newcommand{\lumc}{\textsc{lumc}}

\usepackage{graphicx}
\usepackage{natbib}

\begin{document}
 
\title{Testing Light Unaffiliated Mass Clumps in MACS\,0416 on galaxy and galaxy cluster scales
using JWST}
   \titlerunning{Light Unaffiliated Mass Clumps in MACS\,0416}
   \authorrunning{Limousin et~al.}
   \author{Marceau Limousin\inst{1}, Derek Perera\inst{2}, Gregor Rihtar{\v{s}}i{\v{c}}\inst{3}, Liliya L.~R. Williams\inst{2} \&
Jori Liesenborgs\inst{4}
      \thanks{Based on observations obtained with the \emph{James Webb Space Telescope (JWST)}}
       }
   \offprints{marceau.limousin@lam.fr}

   \institute{
$^1$ Aix Marseille Univ, CNRS, CNES, LAM, Marseille, France.\\
$^2$ School of Physics and Astronomy, University of Minnesota, Minneapolis, MN, 55455, USA.\\
$^3$ Faculty of Mathematics and Physics, Jadranska ulica 19, SI-1000 Ljubljana, Slovenia.\\
$^4$ UHasselt – Flanders Make, Digital Future Lab, Wetenschapspark 2, B-3590, Diepenbeek, Belgium.
              }

   
  \abstract
   {
Light unaffiliated mass clumps (\lumc s), \emph{i.e.} dark matter (DM) components without any stellar 
counterparts, have been reported in strong lensing mass reconstructions of
MACS\,0416, both on galaxy and galaxy cluster scales.
On galaxy cluster scale, the most recent \textsc{Lenstool} parametric mass reconstruction based on 303 
spectroscopically confirmed multiple images features a \lumc\, in the South of the cluster.
On galaxy scale, the most recent \textsc{Grale} non-parametric mass reconstruction based
on 237 multiple images features two \lumc s, M1 and M2.
Given the implications of these findings in the context of structure formation and evolution, 
we test these features parametrically, using the
\textsc{Lenstool} code.
First, we show that a mass model where each large scale DM component introduced in the modelling is 
associated with a stellar counterpart can reproduce the 303 multiple images, removing the need
for any cluster scale \lumc\, in \lens.

We then update the \textsc{Grale} non-parametric mass reconstruction using the 303 multiple images,
finding that one of the two galaxy scale \lumc, M1, is no longer significant, while 
M2 remains.
We test M2 by explicitely including it in our parametric model, at the position and with the
mass inferred 
from our updated \textsc{Grale} model.
We find that the inclusion of this \lumc\, does not improve the global RMS, but mildly
improves locally the RMS for one multiple image located close to M2.
Besides, the preferred mass for M2 corresponds to the lowest mass allowed by the adopted
prior.
If we allow the mass of M2 to reach 0, then \textsc{Lenstool} converges to this null value,
consistently rejecting M2.
We present a detailed comparison of parametric and non-parametric models in the M2 area.
It appears that both approaches show very similar surface mass density at this location, with a
5-6\% difference between the mass maps. The difference is that \textsc{Grale} favors a distinct mass
substructure when \textsc{Lenstool} favors a more diffuse mass distribution.

We have been able to propose a parametric mass model without including any \lumc s, providing
further evidences for DM being associated with light in galaxy clusters.
Finally, further investigations on the mass distribution at the M2 location is relevant.

We release in this paper two new mass models and associated products based on the 303 multiple images that will be hosted at 
the Strong Lensing Cluster Atlas Data Base at Laboratoire d'Astrophysique de Marseille.
   }


   \maketitle

\section{Introduction}

Dark Matter (DM) is an intriguing component that is thought to largely dominate the mass budget in
astrophysical objects, in particular in galaxy clusters. The first indirect evidence for a
"missing mass" is almost 90 years old (Zwicky 1937), but we have no definitive 
statement about its existence and very limited clues on its properties, pieces of evidence for
DM being \emph{indirect} only.
At the galaxy cluster scale, both observations and numerical simulations do support the association 
between DM and light, that is the associated stellar component, in most cases in the form of a bright
galaxy.
Observationally, no cluster scale DM clump without any associated light concentration has been
reliably detected so far.
In hydrodynamical simulations, stars do form in the potential well of DM halos
\citep[\emph{e.g.}][]{Borgani_2004,Yoo_2025}.

However, parametric strong lensing (SL) studies of galaxy clusters sometimes display
"misleading features" \citep[see][for a discussion]{Limousin_2022}, in particular
light unaffiliated mass clumps (\lumc s), \emph{i.e.} DM components without any stellar counterparts.
The purpose of these \lumc s is usually to overcome the limitations of parametric
methods, \emph{e.g.} to account for the complex morphology and 
elongation of the clusters which sometimes cannot be properly captured by simple and
idealized analytical formulae.
Still, the interpretation of these \lumc s are not always properly mentioned or
discussed by the authors, which might be misleading for the readers.

Recently, mass models featuring such \lumc s have been revisited in order to 
see if a DM traced by light (the stellar component) model can reproduce the multiple images.
This was found to sometimes be the case, and sometimes not.
\citet[][L22 hereafter]{Limousin_2022} revisited successfully three parametric mass models featuring \lumc s, presenting models where dark matter is traced by light, 
with the goal to probe the inner shape of the DM density profile.
More recently, \citet{Limousin_2025} were unable to propose a mass model where DM is traced
by light in the merging cluster Abell~370. In order to reach a sub-arcsec precision in the
adjustment of the multiple images, a cluster scale \lumc\, had to be 
introduced in the modelling, as found also in earlier works. \citet{Limousin_2025} interpret this 
finding as being symptomatic of the lack of realism of a parametric
description of the DM distribution in such a complex merging cluster. 

The association between DM clumps and light being central to this paper, we briefly
discuss what "associated" means to us.
Actually we refer to a "loose" association: each DM clump has a well identified
stellar counterpart located at most at a few dozen of kpc from it (which is what 
is allowed by self interacting dark matter scenario, see the discussion by L22),
and the other properties are determined entirely by the lensed images, not the
associated stellar component, usually a bright galaxy.
A tight association would imply that DM clumps have the same centre, ellipticity and
position angle of the associated galaxy, and the mass and core radius would be
related to that of the associated galaxy.

\begin{figure*}
\begin{center}
\includegraphics[scale=1.0,angle=0.0]{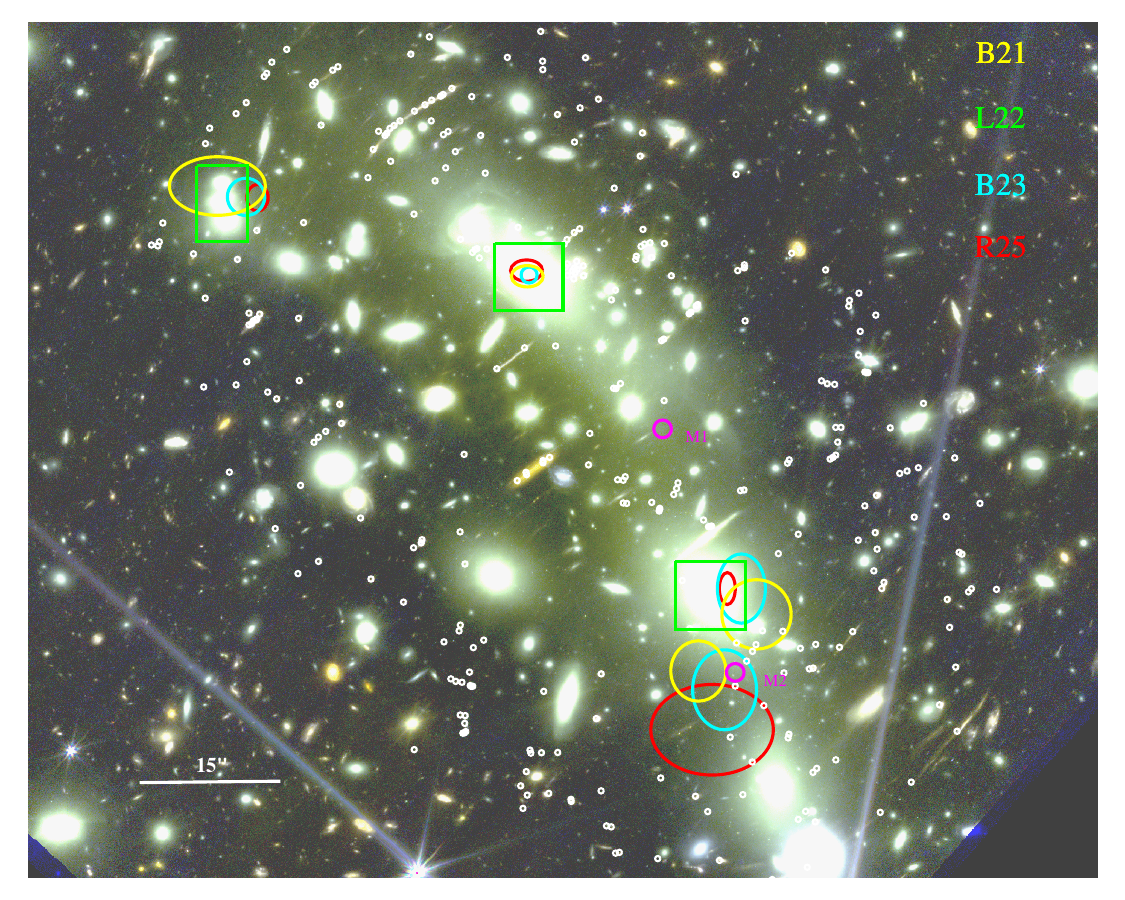}
\caption{Core of \lens\, from \textsc{JWST} data. North is up, East is left.
We show the location of the cluster scale DM clumps inferred in different studies by ellipses whose semi-axis are
equal to the 3$\sigma$ error bars on the position of these clumps:
in yellow for \citet{Bergamini_2021}, in cyan for \citet{Bergamini_2023} and in red for
\citet{Gregor_2025}.
Green boxes represent the prior on the position of each cluster scale DM clumps proposed by \citet{Limousin_2022} and investigated in this work.
We show in magenta the location of the two galaxy scale light unaffiliated mass clumps inferred by
\citet{Perera_2025a}.
In white, the 303 secure multiple images reported by \citet{Gregor_2025}.}
\label{fig1}
\end{center}
\end{figure*}

Galaxy cluster \lens, which has been several times a record holder in terms of number
of multiple images reported \citep{Jauzac_2014, Bergamini_2023, Diego_2024, Gregor_2025}, displays \lumc s, 
both on cluster \citep{Bergamini_2021, Bergamini_2023, Gregor_2025} and galaxy scales 
\citep{Perera_2025a}.
In this paper, we aim to test if we can present a competitive mass model without the need
for such \lumc s, \emph{i.e.} where all (cluster and galaxy scale) DM components have a clear stellar
counterpart.
The light distribution in the core of \lens\, is trimodal (see green boxes on Fig.~\ref{fig1}).
There are two dominant BCGs, and a sub-dominant light concentration
in the north-east.
We use JWST data (F356W, F277W, and F115W filters) presented in \citet{Gregor_2025} in
order to generate Fig.~\ref{fig1} and Fig.~\ref{fig2}.

Besides, the number of haloes constituting a
given galaxy cluster is also important to investigate since 
the abundance of haloes/sub-haloes is a probe of the hierarchical structure formation
scenario \citep[\emph{e.g.}][]{Allen_2011,Jauzac_2015}. 

All our results use the $\Lambda$CDM concordance cosmology with 
$\Omega_{\rm{M}} = 0.3, \Omega_\Lambda = 0.7$, and a Hubble constant 
\textsc{H}$_0 = 70$ km\,s$^{-1}$ Mpc$^{-1}$. 
At the redshift of \lens\, ($z$\,=\,0.396), this cosmology implies a scale of 
5.34\,kpc/$\arcsec$.

\section{\textsc{Lenstool} Parametric mass reconstructions}
A number of parametric mass reconstructions using \textsc{Lenstool} \citep{jullo07} have been performed.
We consider and discuss here only the one performed since 2020, and we present a new one.
They all use secure multiple images with known spectroscopic redshifts.
They incorporate hot-gas mass distribution inferred from X-ray data.
These models describe \lens\, as a superposition of cluster scale and galaxy scale DM
halos, all described using a dPIE mass distribution \citep{Limousin_2005,ardis2218}.
Galaxy scale halos are associated with individual cluster members using the
scaling laws described in these works in order to relate their
mass with their luminosity. Importantly, these scaling laws are constrained using a
prior which is based on the measure of the stellar velocity dispersion of cluster
members \citep[see][for details]{Bergamini_2021, Bergamini_2023}.
Two galaxy scales halos are optimised individually instead of using the scaling laws, since
they were found to have an influence on some multiple images.

Besides, a number of \emph{cluster scale} DM halos are introduced in the modelling.
This is discussed in the following.

\subsection{pre-\emph{JWST} data}

Using Hubble Frontier Fields (HFF) and \textsc{buffalo} imaging data, complemented with
ground based spectroscopy (in particular MUSE), many secured multiply imaged systems 
were found in \lens.

\citet[][B21 hereafter]{Bergamini_2021} reported 182 multiple images.
They are reproduced by a 4 DM clumps model, whose positions are shown as yellow
ellipses on Fig.~\ref{fig1}. 
One of them, in the South, is not affiliated with light.
This model reproduces the multiple images with an RMS of 0.40$\arcsec$.
B21 also presented mass models that reproduce the 182 multiple images fairly well
(RMS between 0.45$\arcsec$ and 0.48$\arcsec$ instead of 0.40$\arcsec$) where the
DM is described using three mass clumps only.
Based on three figures of merit (matching of the internal kinematics of the cluster
members galaxies as well as BIC and AIC criteria), they selected the four mass clumps 
model as their reference.

L22 revisited this mass model by imposing that
each cluster scale DM halo must coincide with a luminous counterpart, imposing the
position of each of the three DM halos to be within the green boxes shown on Fig.~\ref{fig1}.
They reproduced the 182 multiple images with an RMS of 0.49$\arcsec$.

Adding new MUSE spectroscopic data, \citet[][B23 hereafter]{Bergamini_2023} reported 
237 multiple images.
They are reproduced with an RMS equal to 0.43$\arcsec$ by a 4 DM clumps model, whose
positions are shown as cyan ellipses on Fig.~\ref{fig1}.
One of them, in the South, is not affiliated with light.

\subsection{\emph{JWST} data}

\citet{Diego_2024} were the first to examine the \emph{JWST} NIRCam imaging data
from Prime Extragalactic Areas for Reionization and Lensing Science (PEARLS) 
survey, publishing a lens model along with a list of 343 multiple image 
candidates.
Subsequently, \citet[][R25 hereafter]{Gregor_2025}, 
identified additional candidates using NIRCam imaging from the CAnadian NIRISS Unbiased Cluster Survey
(CANUCS) and obtained spectroscopic redshifts for 
a subset of the \citet{Diego_2024} and CANUCS candidates with \emph{JWST} NIRSpec and NIRISS spectroscopy. This led to a final catalog of 303 multiple images with spectroscopic confirmation, including most of the pre-\emph{JWST} systems reported by B23, \citet{Richard_2021}, and \citet{Diego_2024_preJWST}.
These images are reproduced with an RMS equal to 0.53$\arcsec$ by a 4 DM clumps model, whose
positions are shown as red ellipses on Fig.~\ref{fig1}.
One of them, in the South, is not affiliated with light.
Actually, neither B23 nor R25 experimented with alternative parametrisations.
Instead, they adopted the best performing parametrisation from B21, as
determined by statistical
metrics, rather than following the L22 approach of associating each large scale DM
halo with a luminous counterpart.
As a result, the Southern cluster scale \lumc\, was not independently identified by the different
authors; rather, these models are inherently linked through the same prior assumption.

Following L22, we here revisit this mass model by imposing the position of each
of the three DM halos to be within the green boxes (Fig.~\ref{fig1}).
Our aim is to see if, using 303 images instead of 182, we are able to reproduce the
multiple images using a DM traced by light model.
This turns out to be possible, reaching an RMS equal
to 0.57$\arcsec$, \emph{i.e.} comparable with what is found by R25 
(0.53$\arcsec$)\footnote{The RMS difference of 0.04$\arcsec$ is of the same order
of magnitude as the scatter in RMS values found when running the same model of 
\lens\, with
different values of \textsc{rate} and \textsc{Nb} (see Table~\ref{rateNb}).
This difference is an order of magnitude smaller than the typical RMS error
expected from image misidentification \citep{Jauzac_2016} or from unaccounted
structure along the line of sight \citep{host}.
}.
We present and discuss this model in Appendix A and we refer to it as L25 in the following.
In particular, we perform the (\textsc{rate}, \textsc{Nb}) test 
\citep[see][and Appendix~A]{Limousin_2025} in order
to check the convergence of the MCMC chains and the reliability of our results.
We compare how increasing from 182 to 303 images
improves constraints on the DM clumps parameters.

We conclude that there is \emph{no need} for any cluster scale \lumc s in \lens.
We will compare the \emph{total} projected masses from all these models in Section~4.

\section{\textsc{Grale} non-parametric mass reconstructions}

\subsection{pre-JWST data}

\citet[][P25 hereafter]{Perera_2025a}, using the 237 images reported by B23 as constraints, 
presented a free-form mass reconstruction of \lens\, using the lens inversion
algorithm \textsc{Grale} \citep{Jori_2006,Jori_2007,Jori_2020}.
The 237 images are reproduced with an RMS equal to 
0.19$\arcsec$.
They find broad agreement with previous reconstructions, and identify two
$\sim$\,10$^{12}$ M$_{\sun}$ \lumc s (M1 \& M2, see Fig.~\ref{fig1}).
The procedure to identify \lumc\, is based on the following criteria.
First, the background subtracted mass within the clump should be of the order of a galaxy scale 
halo, typically exceeding 10$^9$ M$\sun$ within a few kpc.
Second, the clump should not be associated with any galaxy, i.e. located 
more than 2$\arcsec$ from any visible galaxy.
Finally, the clump should lie sufficiently close to some multiple images (typically 
within 2$\arcsec$).

\subsection{JWST data}

Using the 303 images reported by R25, we present an updated mass model to the one presented in P25
and that we refer to as PCANUCS in the following. The full surface mass density is shown in Figure \ref{gralemodel}. The model is generated in the same way as in P25, so we refer the reader to Section 3.1 of that article for specific details on the free-form modelling process. The updated model reconstructs the 303 observed images with an RMS of 0.23$\arcsec$,
which is comparable to that of the original. In general, the updated model is broadly in agreement with the original model.

The key difference of interest is the presence of M1 and M2. The updated model does not recover M1, with the mass distribution remaining smooth in that region. 
Since there are no new JWST images in the vicinity of M1, 
and the closest original images are 10$\arcsec$ away,
we can adequately confirm that the M1 feature recovered in the original model is likely a shape degeneracy from the underconstrained region rather than a real \lumc. M2, on the other hand, remains in the updated model. In the original model, M2 was recovered at RA = 64.03114, Dec = -24.07990 with a mass of $5.7 \pm 0.2 \times 10^{11} M_{\sun}$ within 1.5$\arcsec$. In the updated model, M2 is recovered in nearly the same position as before (RA = 64.03071, Dec = -24.07974). 
M2 now is less peaked as in the original model, manifesting more as a mass extension from the Southern cluster scale halo. To quantify its parameters, we quote its fiducial characteristic radius as the point of inflection of the circularly averaged profile centered about M2's peak. With this definition, the 
characteristic radius of M2 in the updated model is 2.5$\pm$0.3$\arcsec$, and the mass within this radius 
is $4.4 \pm 1.0 \times 10^{11} M_{\sun}$. 
Note that, given its mass, in a $\Lambda$CDM Universe, under the assumption
of TNG50's galaxy formation
model, M2 should host a luminous counterpart \citep{Doppel_2025}. 
The persistence of M2 in 
the \textsc{Grale} models makes it an interesting candidate for a potential \lumc, and we test this hypothesis in the following.

\section{Comparing Models}

We compare in this Section the mass models that have been derived using the 
303 images, \emph{i.e.} PCANUCS, R25 and L25, which do reproduce the multiple images with an RMS equal to 0.23$\arcsec$, 0.53$\arcsec$ and 0.57$\arcsec$
respectively.

\subsection{Total Projected Mass Distribution}

\begin{figure*}
\begin{center}
\includegraphics[scale=1.0,angle=0.0]{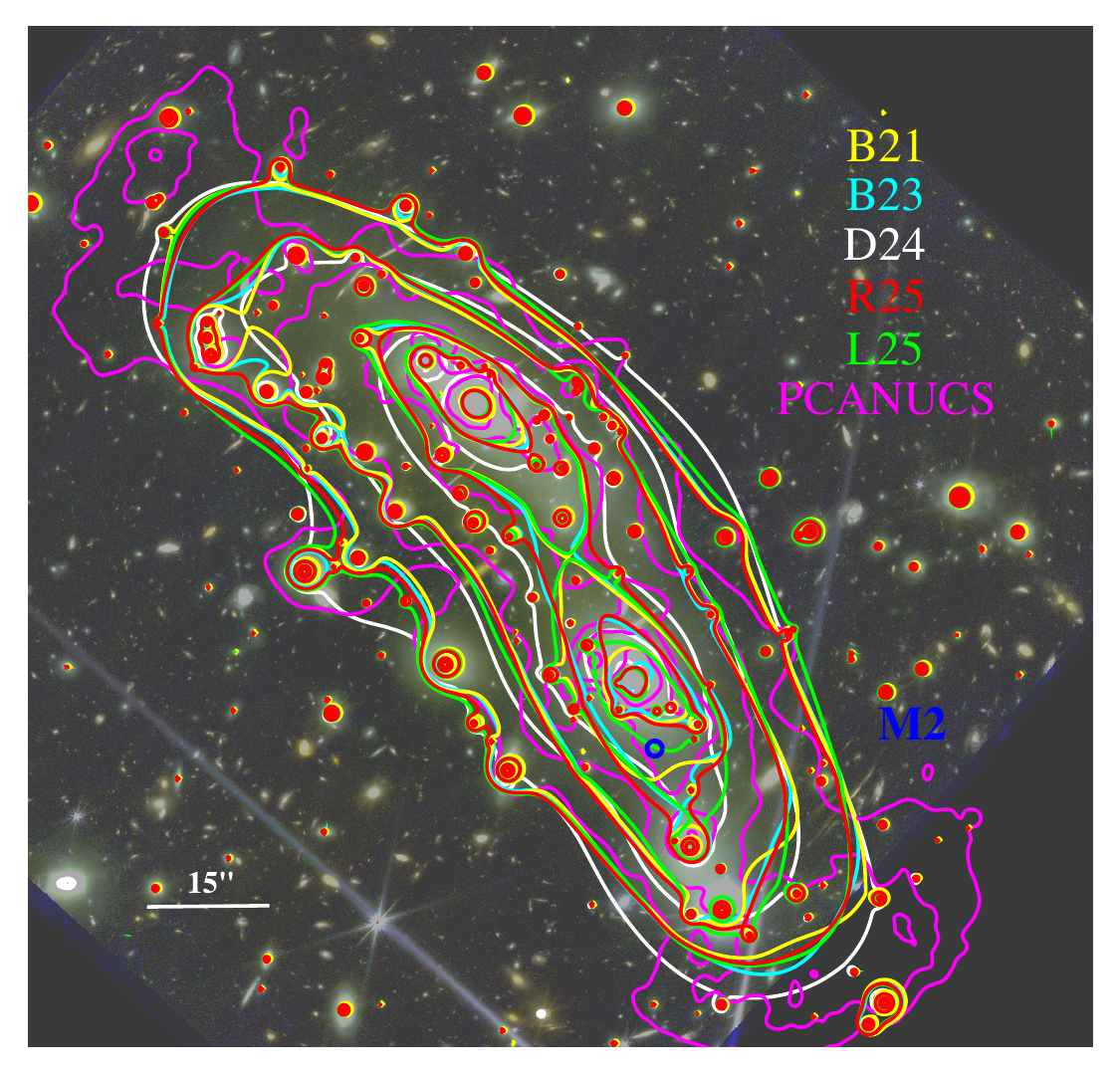}
\caption{Core of \lens\, from \textsc{JWST} data.
We show the \emph{total} convergence maps for all models:
B21 in yellow; B23 in cyan; R25 in red; our updated \textsc{Grale} model
(PCANUCS) in magenta and our
updated \textsc{Lenstool} model (L25) in green.
We also show the convergence map from \citet{Diego_2024} in white (D24).
We show the location of M2 in blue.}
\label{fig2}
\end{center}
\end{figure*}

We now turn to compare the \emph{total} convergence maps of all the 
mass models discussed in this paper, \emph{i.e.} accounting both for the DM and
the galaxy scale components.
We show this comparison on Fig.~\ref{fig2}.
The agreement is very good, in particular between the different \textsc{Lenstool}
reconstructions which are pretty much indistinguishable.
\textsc{Grale}, which does not include cluster galaxies as input,
recovers the existence of several galaxy members, which show up as wiggles in the
density contours.
It features two "wings" in the North east and in the South west,
where no more multiple images are found, implying a large uncertainty in the mass
distribution outside the region confined by the multiple images. 
Usually, most \textsc{Grale} features located outside of the strong
lensing area can be ignored as degeneracies.
We also present on Fig.~\ref{fig2} the convergence map obtained by \citet{Diego_2024} using
\textsc{WSLAP+}, a hybrid mass reconstruction method in which the
surface mass density is modeled as a combination of a soft component describing the
DM and a compact component accounting for the mass associated with individual
cluster galaxies \citep{Diego_2025}.
We find the resulting convergence map to be in very good agreement with the other
reconstructions.

It is interesting to compare Fig.~\ref{fig1} and Fig.~\ref{fig2}.
If the descriptions of the underlying DM
distribution from the parametric \textsc{Lenstool} models do disagree with each other,
in particular the ones by B21, B23 and R25 (displaying four cluster scale DM haloes)
with the one presented in this work, the total convergence (hence mass) from all these
models is in very good agreement.
We stress that the galaxy scale component is here constrained via stellar velocity
dispersion measurement,
which limits the degeneracy between both components \citep[see discussion in][]{Limousin_2016} and is a major step forward in SL modelling.
Still, we do suffer from residual degeneracies which prevent us from discriminating
a four DM clumps model from a three DM clumps mass model, even with 303 multiple
images.
Part of these residual degeneracies might come from the one between the smooth and galaxy scale
components.
Indeed, the scaling laws used to relate cluster member masses to their luminosity
is constrained, not the mass associated with individual galaxies, in particular
the BCGs which definitely have a strong contribution in the area where the
different studies disagree (here in the Southern part of \lens).
The next step might be to constrain individually the mass of the brightest cluster members using spectroscopy, which is within our observing capabilities and our algorithms
\citep{Beauchesne_2024}.

\subsection{Testing M2}

The region South of the Southern BCG is worth studying more closely because 
many models (B21, B23, R25, P25) do place extra mass there, though the 
specific distribution differs between models; see Fig.~\ref{fig1}. 
Most parametric models from the literature prefer a diffuse cored cluster scale 
component, while \textsc{Grale} finds a more compact structure, M2. 
This motivates us to test M2, using \textsc{Lenstool}.

We consider the \textsc{Lenstool} mass model discussed
in Section~2.2 and presented in Appendix A (L25).
We explicitely add a mass clump at the location of M2, whose position can
vary by $\pm$ 1$\arcsec$, and whose mass is equal to
the one derived from our updated \textsc{Grale} mass reconstruction.
We use a circular dPIE profile to describe M2, with a vanishing core radius and a
scale radius set to 40\,kpc (the value of these parameters is found to have no
influence on the results). 
Its velocity dispersion is allowed to vary between 120 and 250 km/s, encompassing
the mass derived from our \textsc{Grale} mass reconstruction.
We optimise this new model (all parameters are optimised)
using the 303 multiple images as constraints.

The RMS is equal to 0.58$\arcsec$, equal to the RMS found in Section~2.2 when
M2 is not included. Besides, the optimised value for M2's velocity dispersion
is equal to 120 km/s, \emph{i.e.} the lower bound allowed by the prior.
We refer to this model as Model\,1 in the following.
If we then allow the velocity dispersion of M2 (hence its mass) to reach 
0 and rerun this model, the optimised value for M2's velocity dispersion
is equal to 0 km/s. The RMS stays stable to 0.58$\arcsec$.
We refer to this model as Model\,2 in the following.
Therefore, it appears that \textsc{Lenstool} wants to remove M2. 

If the total RMS is not improved by the inclusion of M2, we consider the 
individual RMS for the three multiple images that are closest to M2, 
namely C2.3, K29.2 and K45.2, following the CANUCS IDs.
Let's call Model\,0 the mass model presented in Section~2.2, where M2 is not 
included.
RMS$_i$ corresponds to the individual RMS for a given image corresponding
to Model\,$i$.
We report in Table~\ref{tableRMS} the individual RMS for these three images, 
together with their distance to M2.

Since in Model\,2, M2 is removed (its velocity dispersion tends to 0), we
can consider that Model\,0 and Model\,2 are essentially the same.
Therefore we can use the difference of RMS between Model\,0 and Model\,2
as a measure of the fluctuation in the modelling process, and compare it to
the difference of RMS between Model\,0 and Model\,1 to see if it is significant.
This fluctuation vary between 0.02$\arcsec$ and 0.07$\arcsec$, depending on which of
the three images we consider.
To be conservative, we set this fluctuation to 0.1$\arcsec$, which is of the same order
of magnitude as the scatter from the posterior sample.
We report in Table~\ref{tableRMS} $\Delta$(RMS), the difference between RMS$_0$ and RMS$_1$.
It is found to be equal to 0.13, 0.32 and 0.01$\arcsec$ for images
K45.2, C2.3 and K29.2 respectively. 
For each image, we find that $\Delta$(RMS) is positive, which means that adding 
M2 does locally improve the RMS of the
considered image. This improvement is much smaller than the fluctuation
for image K29.2;
it is of the order of the fluctuation for image K45.2, and 
equal to three times the fluctuation for image C2.3.
We conclude that this improvement is significant for one image out of the three
closest images to M2.

\begin{table*}
\begin{center}
\begin{tabular}{ccccccc}
\hline \\*[-1mm]
Image & $d$ & RMS$_0$ & RMS$_1$ & RMS$_2$ & $\Delta$(RMS) & PCANUCS$_{\rm RMS}$\\
\hline \\*[-1mm]
K45.2 & 1.54 & 0.78 & 0.65 & 0.75 & 0.13 & 0.03\\
\hline \\*[-1mm]
\smallskip
C2.3 & 1.86 & 0.45 & 0.13 & 0.43 & 0.32 & 0.14\\
\hline \\*[-1mm]
\smallskip
K29.2 & 3.06 & 0.15 & 0.14 & 0.22 & 0.01 & 0.05\\
\hline \\*[-1mm]
\smallskip
\end{tabular}
\end{center}
\caption{Information regarding the three multiple images that are closest to 
M2: CANUCS ID; distance $d$ to M2 in arcsec; individual RMS (distance between observed
and reconstructed images, in arcsec) 
for each image for the different
models investigated: 0 corresponds to the model presented in Section~2.2;
1 to the model explicitly including M2 with a mass derived from the \textsc{Grale} updated analysis; 2 to the model where the mass of M2 can reach 0.
We report $\Delta$(RMS), the difference between RMS$_0$ and RMS$_1$,
as well as PCANUCS$_{\rm RMS}$, the RMS derived from our updated
\textsc{Grale} model.}
\label{tableRMS}
\end{table*}

To test M2 non-parametrically as well, we removed the three multiple images closest to M2 and reran a 
\textsc{Grale} model.
Despite this, \textsc{Grale} still favors a mass feature at M2's location with
essentially the same characteristics as before. The total RMS for this model is
0.20$\arcsec$, 
very similar to the previous one (0.23$\arcsec$).
This gives us an estimate of the fluctuation from modelling for \textsc{Grale}
equal to $\sim$0.03$\arcsec$.

\subsection{Model Comparison near M2}

Since the M2 feature persists in reconstructions using \textsc{Grale} and not 
\textsc{Lenstool}, it is useful to quantify the similarity between models in this region, which
is important for the interpretation of M2. 
To compare models around M2, we measure the similarity of the surface mass 
density distribution $\kappa$ as a function of the distance from M2. 
This allows us to see if the disagreement between the models is restricted to positions adjacent to M2. We use the Median Percent Difference (MPD) metric to quantify the similarity in the M2 region (Perera et. al. 2025b). The MPD metric is defined as the median of the percent difference map calculated between models $\kappa_1$ and $\kappa_2$ at grid points $\boldsymbol{\theta}$:
\begin{equation}
    {\rm MPD} = \text{med}\left \{ 2\frac{|\kappa_1(\boldsymbol{\theta}) - \kappa_2(\boldsymbol{\theta})|}{\kappa_1(\boldsymbol{\theta}) + \kappa_2(\boldsymbol{\theta})} \times 100 \right \}
\end{equation}\label{eq:PD}
We choose this metric over the other two used in \citet{Perera_2025b} since the MPD is the most resistant to the map resolution, which is important for such a small scale. The smaller the value of MPD, the more similar two models are to one another.

\begin{figure*}[h!]
\begin{center}
\includegraphics[scale=0.83,angle=0.0]{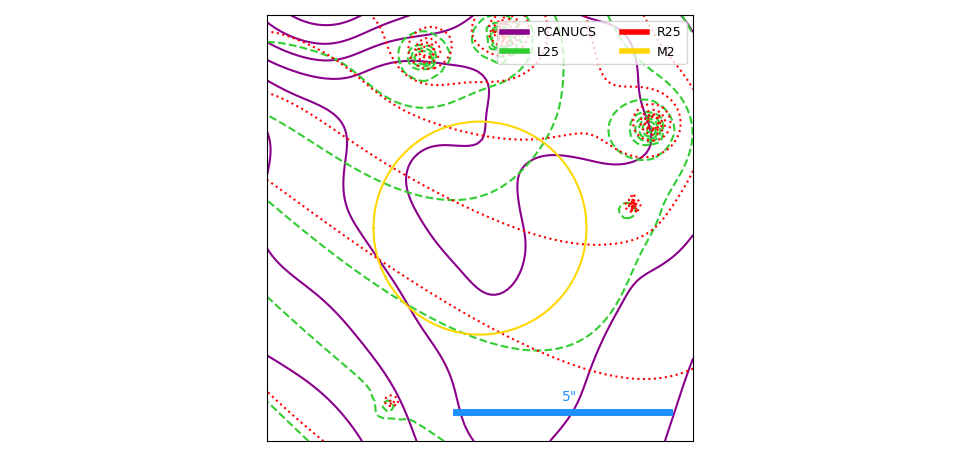}
\includegraphics[scale=0.77,angle=0.0]{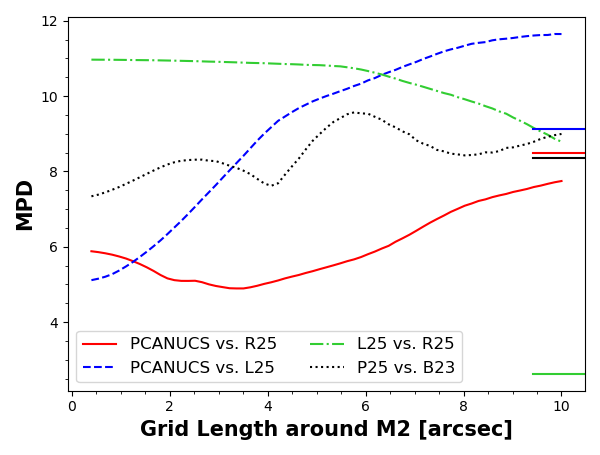}
\caption{{\it Top:} Density contours in the $10^" \times 10^"$ area centered on the peak of the M2 substructure recovered at  RA = 64.03071, Dec = -24.07974 in PCANUCS. Spacing between contours is $\Delta \log_{10}\kappa = 0.05$. The three primary lens models discussed in this article using the 303 images reported by R25 are shown in solid purple (PCANUCS), dashed green (L25), and dotted red (R25). The solid yellow circle denotes M2 with radius of 2.5" corresponding to its characteristic radius. {\it Bottom:} The Median Percent Difference (MPD) between models as a function of the length of the grid comparison window centered about M2. Comparisons between PCANUCS and R25 (solid red), L25 and R25 (dash-dotted green), and PCANUCS and L25 (dashed blue) are shown. To compare these with the earlier works, the comparison using 237 images between P25 and B23 (dotted black) is also shown. For the comparisons between \textsc{Grale} and \textsc{Lenstool} models, the solid horizontal dashes on the right indicate the MPD over the whole cluster for PCANUCS and R25 (red), PCANUCS and L25 (blue), L25 and R25 (green), and P25 and B23 (black). We note that the apparent misalignment of the contours near the galaxy positions is a likely side effect of interpolation using models of varying resolution. These shifts are restricted to the pixel level, and have a negligible effect on our results.}
\label{figM2}
\end{center}
\end{figure*}

We calculate the MPD between models on square grids of steadily increasing side length. Each grid is centered on the peak of M2, and each model is interpolated onto the same grid for comparison. We establish the upper limit on the grid length to be 10", as beyond this limit the mass of the BCG begins to dominate. This comparison is repeated for the three models discussed in this paper using 303 images for reconstruction: L25, R25, and PCANUCS. This gives us a total of 3 comparisons as a function of grid length. These 3 mass models are shown in the top panel of Fig.~\ref{figM2} up to the 10" grid limit. 
The bottom panel of Fig.~\ref{figM2} shows the MPD as a function of the grid length around M2. To include the pre-CANUCS models, we also show the comparison between P25 and B23.
We find that on the smallest scales ($\lesssim2^"$) PCANUCS agrees with L25 and R25 at an MPD $\sim5-6\%$. It is necessary to evaluate the MPD in the context of the entire cluster. For PCANUCS and R25, MPD $= 8.49\%$, and for PCANUCS and L25, MPD $= 9.13\%$. This allows us to conclude that PCANUCS and R25 agree better in the region up to 10" from M2 than over the entire cluster, while PCANUCS and L25 agree better up to 3.6" from M2 than over the entire cluster. This is an improvement from the comparison between P25 and B23, where the MPD on the smallest scales is 7.3\%, and the two models have a MPD of 8.36\% over the entire cluster. Therefore, we conclude that the increase in images from 237 to 303 has led to a closer agreement between \textsc{Grale} and 
\textsc{Lenstool} in the vicinity of M2. 

Interestingly, L25 and R25 disagree substantially at MPD $\sim11\%$ up to 6" from M2. This is significant because L25 and R25 both use \textsc{Lenstool}, and agree over the whole cluster with MPD of 
2.63\%. 
This is explained by the amplitude of the mass density reconstructed in both models. In the M2 region, the value of $\kappa$ is lowest in R25 and highest in L25, with PCANUCS recovered between the two.
The disagreement between L25 and R25 appears to be restricted to this region around M2, as the MPD substantially reduces on scales $>10^"$ to 
$\sim8\%$. 
This local disagreement is likely to be due to the different prescriptions of both models in this area
(Fig.~\ref{fig1}). M2 is located in between the two large scale DM clumps proposed by R25,
whereas L25 assumes a single large scale DM clump coincident with the BCG.
In fact, L25 and R25 agree better on larger scales than PCANUCS does with either \textsc{Lenstool} model, which is consistent with the results of Perera et. al. (2025b), which finds that parametric models are generally more similar to one another on cluster scales. 

Up to the 10" limit of the comparison around M2, we can conclude that R25 favors M2 more than L25 does since it exhibits consistent agreement better than the entire cluster over the entire 10" grid area. That being said, neither \textsc{Lenstool} model reproduces a mass feature similar to PCANUCS. As has been shown, 
\textsc{Lenstool} always seems to prefer removing such a compact mass 
component in this region. The takeaway from this comparison experiment, therefore, is that the surface mass density around M2 exhibits agreement between 
\textsc{Grale} and \textsc{Lenstool} at MPD $\sim5-6\%$, which is better than over the entire cluster. However, the concentration of the density in this region remains unconstrained, with \textsc{Grale} favoring a distinct mass substructure, and \textsc{Lenstool} favoring a more diffuse mass distribution.

In other words, it appears that \textsc{Lenstool} wants to remove an explicit mass component at the M2
location, but that its surface mass density already features some mass at this location whose amount is 
close (within 5-6\%) to the \textsc{Grale} reconstruction.
Therefore, while M2 is not favoured by the \textsc{Lenstool} parametric approach, it is
not ruled out by it.

Note that we have not included the reconstruction by \citet{Diego_2024} in the model
comparison near M2, as it relies on different inputs than those used in R25, L25, 
and PCANUCS, which could
introduce a bias in the comparison.
In particular, both the catalog of multiple images and the catalog of cluster
members differ.

On the day before the formal acceptance of this paper, another study on \lens\, 
was submitted for publication \citep{Cha_Jee_2025}. The authors present a new hybrid
SL modeling algorithm, \textsc{MrMARTIAN}, which they successfully test on simulated 
data and apply to \lens,
using the 412 multiple images published by R25. This catalog includes the 303 
spectroscopically 
confirmed multiple images together with additional candidates.
The convergence map from their reconstruction shows no indication of an
overdensity at the M2 location \citep[][private communication]{Cha_Jee_2025}.
A more detailed model comparison in the vicinity of M2 is not pursued here, given the 
timing of the publications and the differences in the sets of multiple images employed.

\section{Conclusion}

On galaxy cluster scales, we have shown that we are able to describe the DM
distribution in \lens\, with a DM traced by light model.
We conclude that there is no need for any galaxy cluster scale \lumc\,
in \lens, since our model reproduces image positions just as well as the model
with a \lumc\,(RMS of 0.57$\arcsec$ vs. 0.53$\arcsec$).
This was already the case when using 182 multiple images (L22).
We confirm this finding using 303 multiple images.
Note that we do not demonstrate that galaxy cluster scale \textsc{lumc}s do not exist in \lens, 
we only show that we do not need
any to reproduce the multiple images, and we argue that this solution is more in tune with the
theoretical expectations of structure formation and evolution.

On galaxy scales, we have shown that when considering 303 multiple images as
constraints, a \textsc{Grale} reconstruction still favors one \lumc\, instead
of two when "only" 237 multiple images were used.
We have been testing this component parametrically with \textsc{Lenstool}, 
finding a mild
local improvement of the RMS when including explicitly M2 in the modeling.
Overall, \textsc{Lenstool} consistently wants to reject M2, 
whereas \textsc{Grale} consistently favors M2.
Still, we find that \textsc{Lenstool} models do also feature some mass in the M2
region whose amount is close to the \textsc{Grale} one (Fig.~3).
Overall, we have shown that lens model degeneracies remain important both on cluster and 
galaxy scales, despite the exquisite quality of the data.

We note that the intra cluster light component, which is not explicitly taken into account
in this work, is known to be a diffuse component \citep[see, \emph{e.g.}][]{Arnaboldi_2022,Montes_2022}
distributed on scales larger than that of individual galaxies. It should therefore not be responsible for the M2 mass feature.

To conclude, we are able to propose a competitive parametric mass model without the need for any
\textsc{lumc} in \lens.
We also present an updated \textsc{Grale} model.
These mass models and associated products are made publicly available for download at the
Strong Lensing Cluster Atlas Data Base, which is hosted at Laboratoire d'Astrophysique de
Marseille\footnote{https://data.lam.fr/sl-cluster-atlas/home.}.
This work provides further
evidence for DM being associated with light at galaxy and galaxy
cluster scales.
This association is loose, as discussed in the introduction.
Finally, further investigations on the mass distribution at the M2 location is relevant.

\section*{Acknowledgment}

We thank the referee for a constructive report, and Jos\'e Maria Diego for kindly sharing his
$\kappa$ map with us.
ML acknowledges the Centre National de la Recherche Scientifique (CNRS) and the
Centre National des Etudes Spatiale (CNES) for support.
This work was performed using facilities offered by CeSAM (Centre de donnéeS Astrophysique de Marseille).
Centre de Calcul Intensif d’Aix-Marseille is acknowledged for granting access to its high performance computing resources.
DP acknowledges the computational resources provided by the Minnesota Supercomputing Institute, which were critical for this work.
GR acknowledges support from the ERC Grant FIRSTLIGHT, the Slovenian national research agency
ARIS through grants N1-0238 and P1-0188 and the European Space Agency through Prodex Experiment
Arrangement No. 4000146646.
This work is based on observations made with the NASA/ESA/CSA James Webb Space Telescope. 
The data were obtained from the Mikulski Archive for Space Telescopes at the Space Telescope 
Science Institute, which is operated by the Association of Universities for Research in 
Astronomy, Inc., under NASA contract NAS 5-03127 for JWST. 
These observations are associated with programs 1176 and 2738 
\citep[PEARLS][]{PEARLS} and program 1208 \citep[CANUCS, DOI: 10.17909/18nv-np70][]{CANUCS}.

\bibliographystyle{aa} 
\bibliography{draft}

\begin{appendix}
\section{\textsc{Lenstool} Mass Model using 303 multiple images.}

The updated \textsc{Lenstool} mass model presented in this paper is very similar to
former \textsc{Lenstool} models discussed in this work, allowing easy comparison
of their respective results.
The cluster members are described in the modelling using scaling laws which
have been constrained by spectroscopic observations (see B23 for more details). 
Two galaxies are optimized individually instead: one cluster member located near the
northern BCG which is responsible for producing additional multiple images locally, 
and a foreground galaxy that perturbs the Southern giant arc.
They are located at 47$\arcsec$ and 23$\arcsec$ from M2, respectively, and therefore 
have a negligible impact on M2\footnote{The optimized velocity dispersion for the 
foreground galaxy reported by R25 and L25 is 136$\pm$27 and 193$\pm$20 km/s (1$\sigma$ error bars).
Assuming a velocity dispersion equal to 165\,km/s for this galaxy, it would generate a convergence
$\kappa$ of 0.006 at the location of M2. This is negligible compared to the reported 
$\kappa$ value of 1.3 at M2, as evaluated from the optimized mass model and shown in Fig.~2.
Considering the deflection angles is also relevant to demonstrate that the gravitational influence
of this foreground galaxy is negligible. 
At the location and redshift of multiple image C2.3, it produces a deflection angle of 0.4$\arcsec$, which is smaller than the total RMS and negligible compared to the total deflection field of $\sim 16\arcsec$ generated by the entire parametric model.}.
The X-ray gas component is explicitely taken into account by incorporating four dPIE
mass clumps, which are not optimised, and whose values have been constrained by
\citet{Bonamigo_2018}.
The main difference between these mass models resides in the number of large scale DM
clumps introduced in the modelling.
The new one presented here and in L22 both do require that each large scale
DM clump must be associated with a luminous counterpart, which sets the number of
such clumps equal to three, whereas other studies propose four cluster scale DM mass
clumps.
In L22, contrary to the other studies discussed here, the ellipticity of each 
large scale DM clump was forced to be smaller than
0.7, motivated by the results from numerical simulations found
for \emph{unimodal} clusters \citep{Despali_2016}.
We here release this constraint, since \lens\, is far from being unimodal.

Two key parameters matter when it comes to the convergence of the \textsc{Lenstool} MCMC sampler
\citep{jullo07}: the \textsc{rate} parameter, associated to the burnin phase,
and the number of iterations (\textsc{Nb}), associated to the
sampling phase.
The smaller the rate, the more the sampler will move slowly to the high-likelihood areas and will be less prone to miss a mode of the posterior.
The larger \textsc{Nb}, the larger the number of iterations of the MCMC chains, the best the
parameter space can be sampled.
We refer the reader to \citep{Limousin_2025} for more details.

\begin{figure*}
\begin{center}
\includegraphics[scale=0.38,angle=0.0]{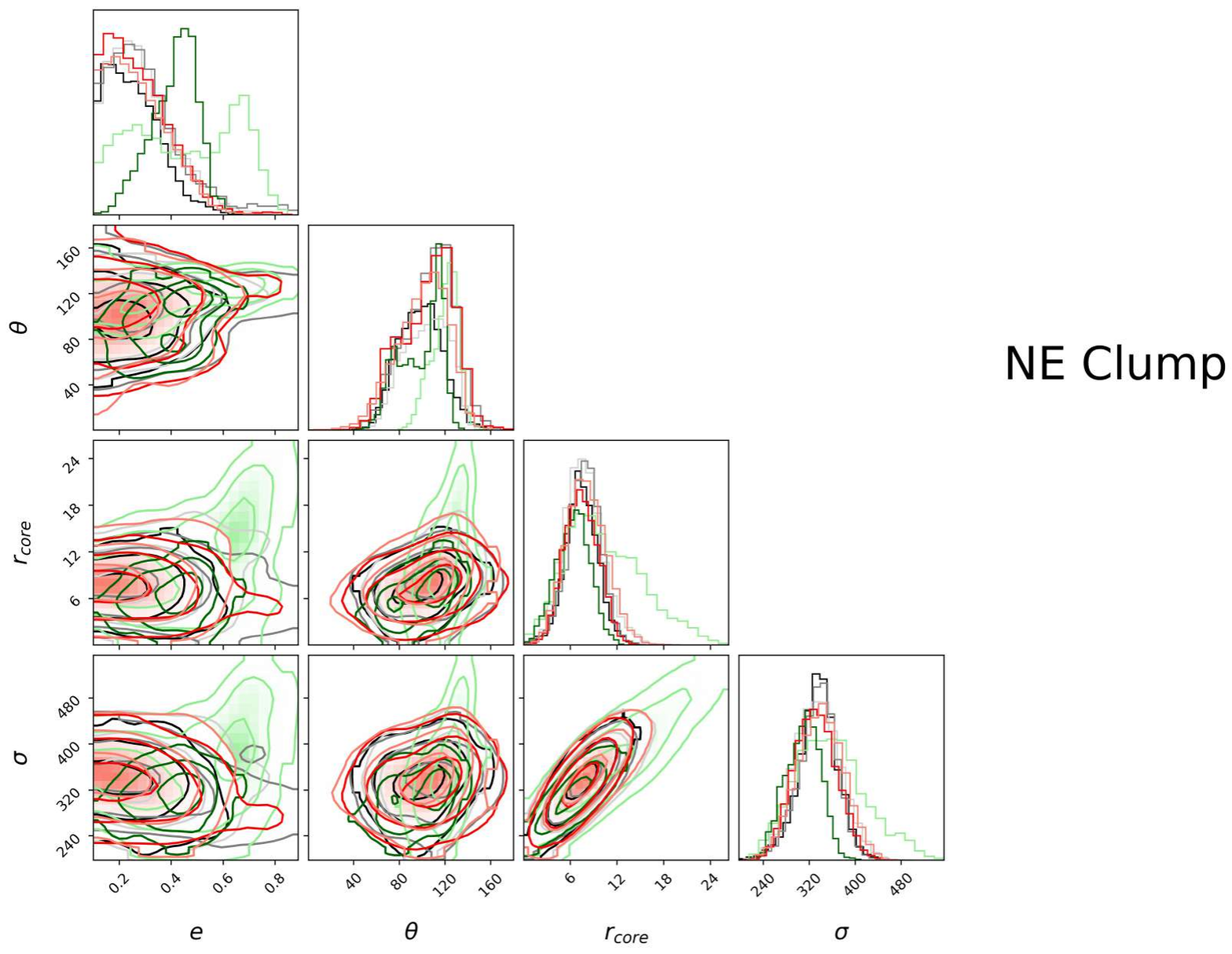}
\includegraphics[scale=0.38,angle=0.0]{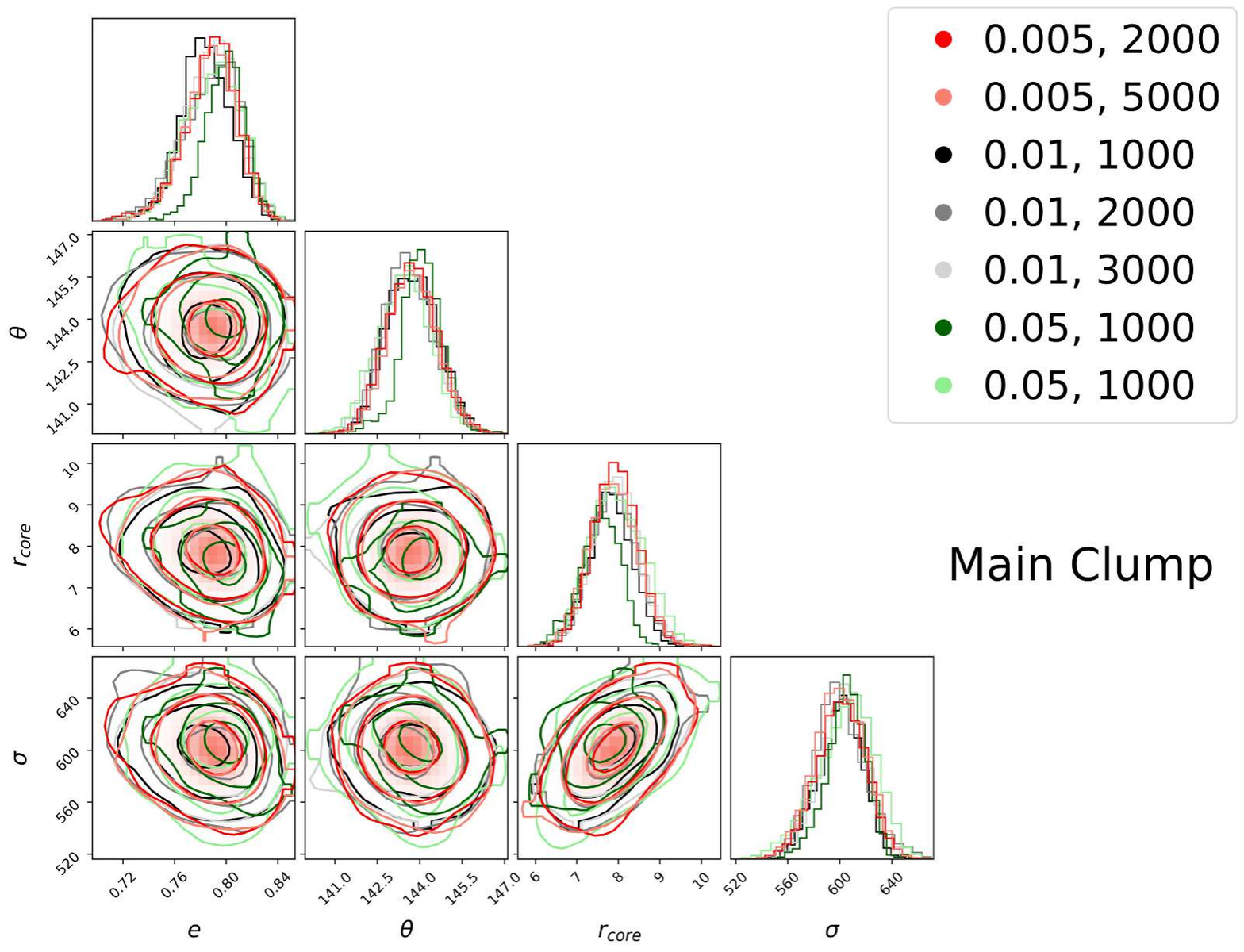}
\includegraphics[scale=0.38,angle=0.0]{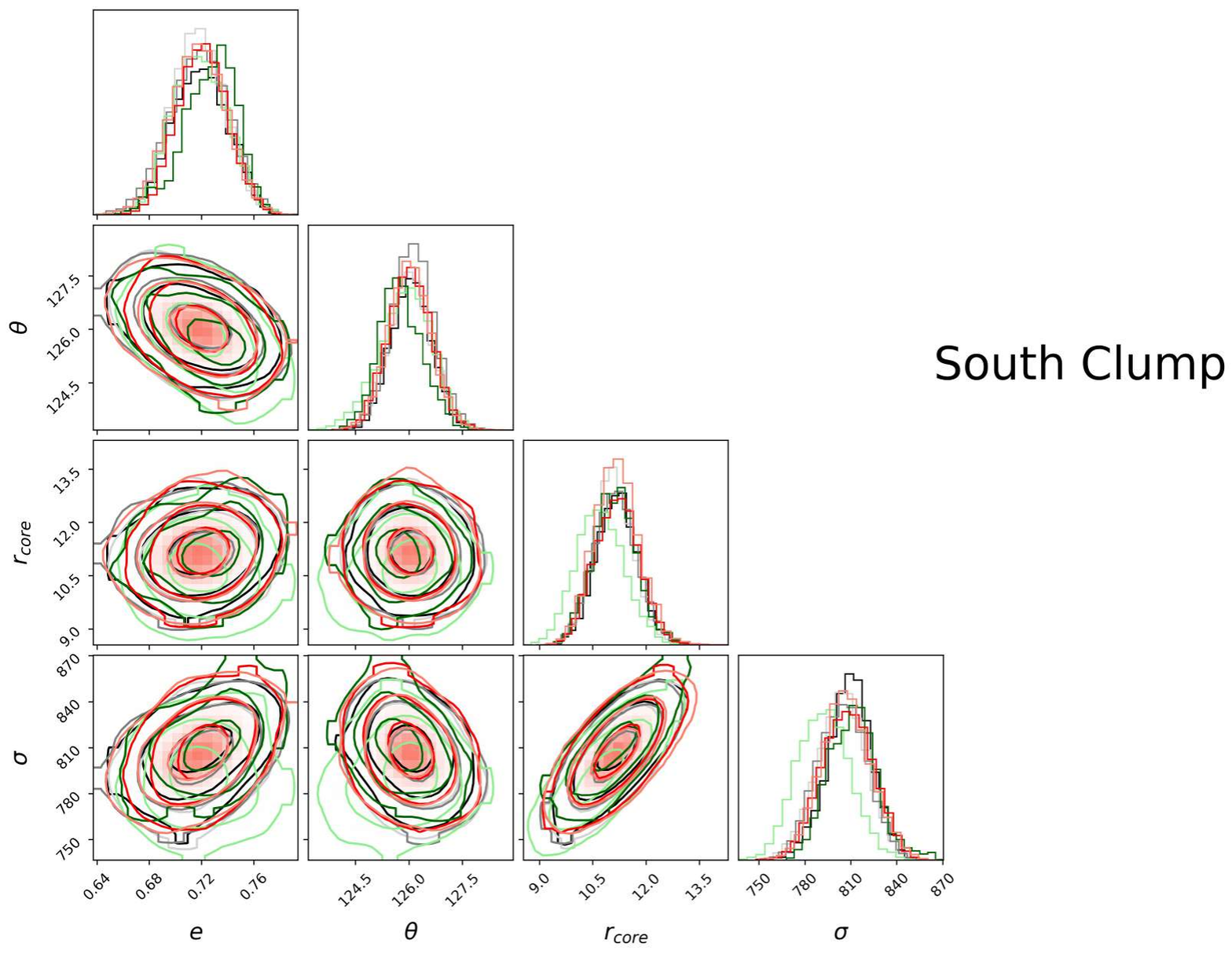}
\caption{Corner plots obtained for the parameters of the mass model, for the values of
\textsc{rate} \& \textsc{Nb} indicated on the legend.\\ \emph{Top:} NE clump; 
\emph{middle}: Main clump;
\emph{bottom}: South clump.
The position of each clump is not shown for clarity since it is constrained to be
coincident with the associated light component.}
\label{fig_rateNb}
\end{center}
\end{figure*}

Former studies have used the following values for (\textsc{rate} \& \textsc{Nb}):
(0.05, 1000) in L22; (0.05, 100000) in B21 and B23 and (0.015, 10000) in R25.
For our updated \textsc{Lenstool} mass model, we perform the 
(\textsc{rate}, \textsc{Nb}) test which was found to be useful to check if the parameters
describing the mass distribution have actually converged \citep{Limousin_2025}.
In short, we lower the \textsc{rate} and increase \textsc{Nb} to see how it influences the
RMS and the parameters of the mass clumps describing \lens.
Convergence is attained when the values for \textsc{rate} \& \textsc{Nb} do not have any influence
on the resulting RMS and on the parameters of the mass clumps.
Table~\ref{rateNb} and Fig.~\ref{fig_rateNb} show the results.
We find that a \textsc{rate} equal to 0.01 is small enough: PDFs corresponding
to the different runs with \textsc{rate} equal to 0.01 (black, grey and light grey)
do agree with each other.
This is not the case for runs with \textsc{rate} equal to 0.05 (dark and light green),
particularly for the NE clump, and, to a lesser extent, the South clump, while the Main
clump is already well defined with a \textsc{rate} equal to 0.05.
Besides, when lowering the \textsc{rate} to 0.005 (red/salmon), the results are in agreement
with the one obtained with a \textsc{rate} equal to 0.01. 
This suggest that, with \textsc{rate} set to 0.01, the models have reached convergence.
Regarding the \textsc{Nb} parameter, we find that \textsc{Nb} equal to 2000 (corresponding to 20\,000 lines in 
the \emph{bayes.dat} file) provides enough sampling. 
Comparing the different runs suggest that, when the \textsc{rate} parameter is small 
enough, hence the area of the parameter space well defined, \textsc{Nb} is of least importance, assuming it is
large enough to provide a decent number of MCMC chains. 2000 is enough in this case.
Our best model, in term of \textsc{rate} \& \textsc{Nb}, is the one with \textsc{rate} equal 
to 0.005 and \textsc{Nb} equal to 5000. It is used to generate the different
quantities presented in this paper. Its parameters are presented in Table~\ref{bestmodel}.

\begin{table}
\begin{center}
\begin{tabular}{ccc}
\hline \\*[-1mm]
\textsc{rate} & \textsc{Nb} & RMS ($\arcsec$) \\
\hline \\*[-1mm]
0.05 & 1000 & 0.59 \\
\hline \\*[-1mm]
0.05 & 1000 & 0.58 \\
\hline \\*[-1mm]
0.01 & 1000 & 0.58 \\
\hline \\*[-1mm]
0.01 & 2000 & 0.58 \\
\hline \\*[-1mm]
0.01 & 3000 & 0.58 \\
\hline \\*[-1mm]
0.005 & 2000 & 0.57 \\
\hline \\*[-1mm]
0.005 & 5000 & 0.57 \\
\hline \\*[-1mm]
\smallskip
\end{tabular}
\end{center}
\caption{RMS obtained given different values of \textsc{rate} \& \textsc{Nb}.}
\label{rateNb}
\end{table}

\input{bestfitable.tex}

\begin{figure*}
\begin{center}
\includegraphics[scale=0.38,angle=0.0]{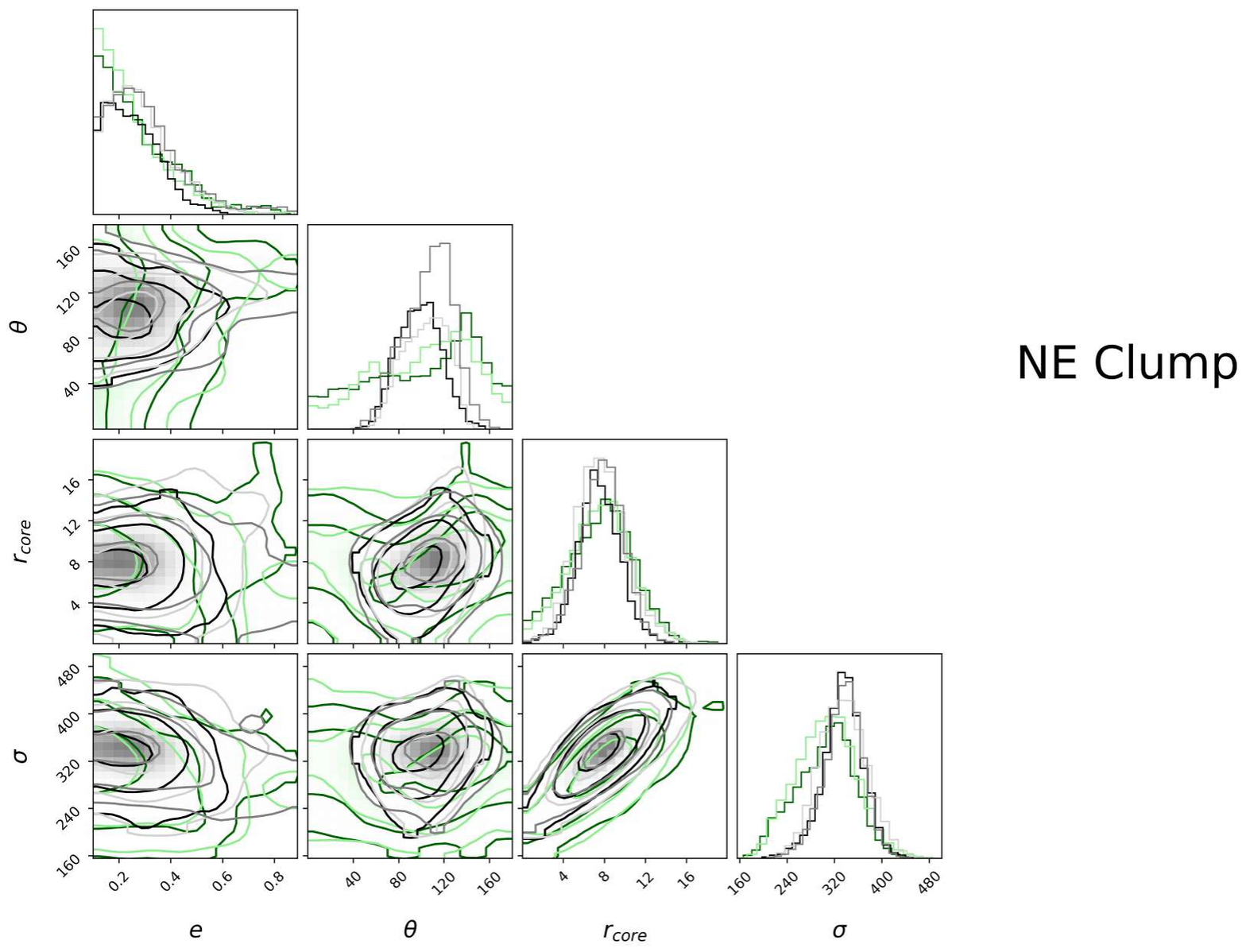}
\includegraphics[scale=0.38,angle=0.0]{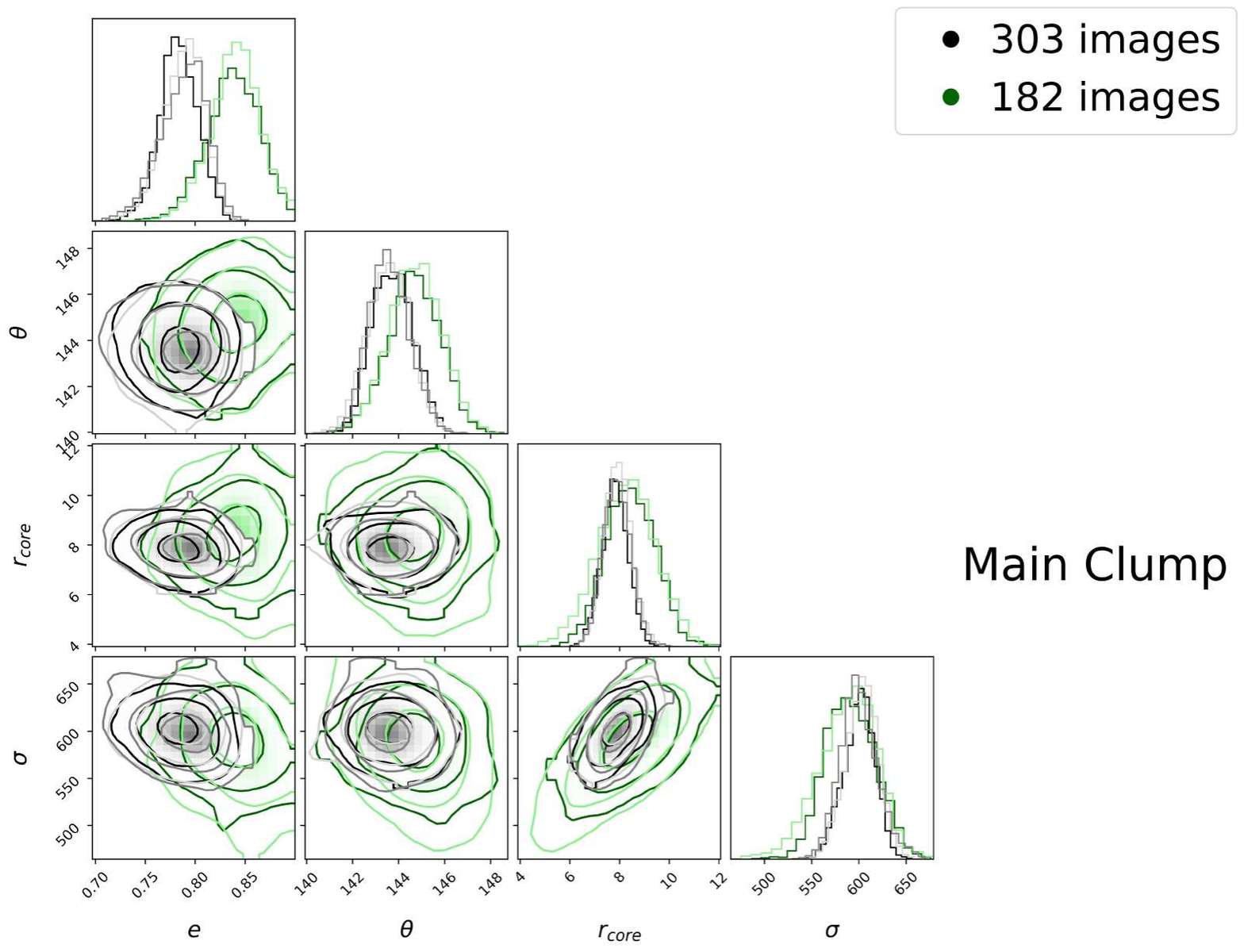}
\includegraphics[scale=0.38,angle=0.0]{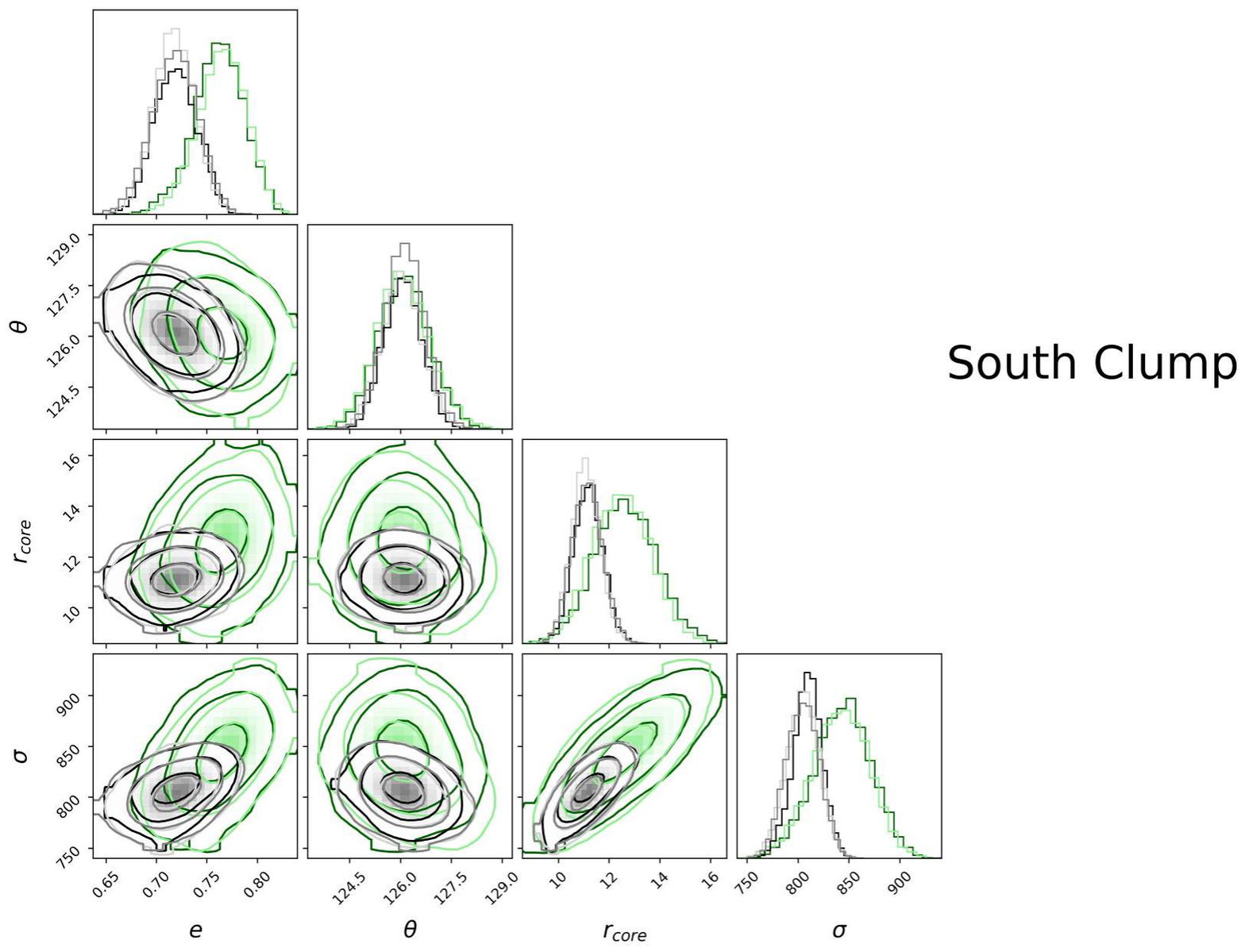}
\caption{Corner plots obtained for the parameters of the mass model when considering
182 multiple images (green) and 303 multiple images (black/grey).
A \textsc{rate} of 0.01 is used for all realisations.}
\label{182vs303}
\end{center}
\end{figure*}

We also compare the parameters obtained on the three DM mass clumps when the number
of multiple images vary from 182 to 303.
Green PDFs on Fig.~\ref{182vs303} correspond to models obtained 
when considering 182 multiple images, with:
(\textsc{rate}, \textsc{Nb}) = (0.01, 1000; dark green) and 
(\textsc{rate}, \textsc{Nb}) = (0.01, 2000; light green). 
In both cases, the RMS is equal to 0.50$\arcsec$.
Grey/black PDFs corresponds to models obtained when considering 303 multiple images,
with 
(\textsc{rate}, \textsc{Nb}) = (0.01, 1000): black,  
(\textsc{rate}, \textsc{Nb}) = (0.01, 2000): grey, and 
(\textsc{rate}, \textsc{Nb}) = (0.01, 3000): light grey.  
In all cases, the RMS is equal to 0.58$\arcsec$.

We can appreciate how the constraints evolve when moving from 182 to 303 images.
PDFs are tighter with 303 images and their position can vary a bit.
Regarding the north-east Clump, we see that with 303 images, its parameters are much
better defined than with 182 images.

\clearpage

\section{\textsc{Grale} Mass Model using 303 multiple images.}

\begin{figure}
\centering
\includegraphics[scale=0.9, angle=0.0]{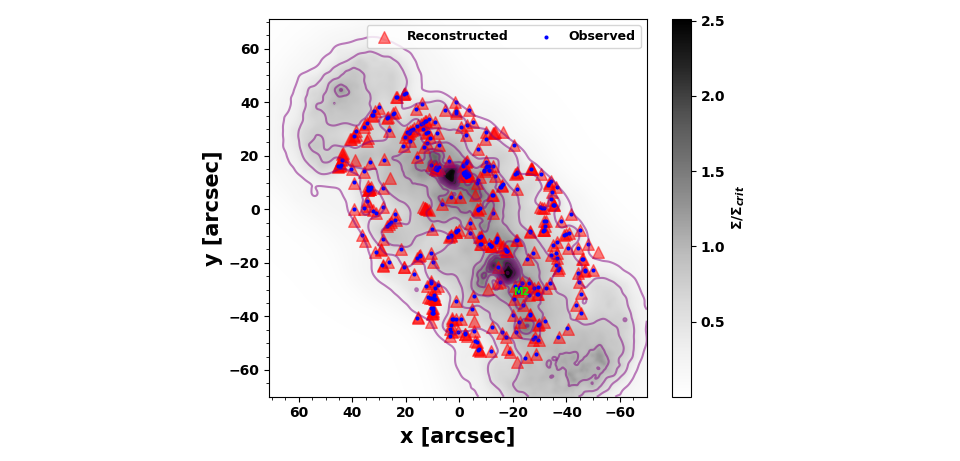}
\caption{Projected surface mass density for the updated \textsc{Grale} model. The observed image positions from R25 are shown as blue dots, while the reconstructed image positions are shown as red triangles. Only the M2 \lumc\, from the original model is reproduced and labelled in green. Contour lines are separated by $\Delta \kappa = 0.25$ in surface mass density.}
\label{gralemodel}
\end{figure}

We present on Fig.~\ref{gralemodel} the projected surface mass density
of our updated \textsc{Grale} model obtained when using 303 multiple images
as constraints.

\end{appendix}

\end{document}

%% file: bestfitable.tex
\begin{table*}
\begin{center}
\begin{tabular}{ccccccc}
\hline \\*[-1mm]
 & $\Delta$\,\textsc{ra} & $\Delta$\,\textsc{dec} &  $e$  & $\theta$ & $\sigma$ (km\,s$^{-1}$) & $r_{\rm core}$ (kpc) \\
\hline \\*[-1mm]
Main Clump &  -2.6$\pm$0.7 (-1.8) & 1.7$\pm$0.5 (1.2)  & 0.78$\pm$0.02 (0.76)  & 144$\pm$1 (143) &  600$\pm$19 (612)  & 42.2$\pm$3.2 (42.7)   \\   
\hline \\*[-1mm]
South Clump  &  22.0$\pm$0.4 (22.3)  & -39.1$\pm$0.6 (-39.4)  & 0.72$\pm$0.02 (0.70) & 126$\pm$1 (126) & 806$\pm$16 (795) &  59.8$\pm$3.2 (57.9)  \\ 
\hline \\*[-1mm]
NE Clump  & -32.6$\pm$1.4 (-31.7) & 9.0$\pm$1.4 (9.0)  & 0.25$\pm$0.1 (0.24) & 104$\pm$21 (99) & 338$\pm$40 (358) & 42.2$\pm$13.3 (45.8)  \\
\hline \\*[-1mm]
\end{tabular}
\end{center}
\caption{dPIE parameters inferred for our best model of MACS\,0416, with an RMS equal to 0.57$\arcsec$, 
obtained with (\textsc{rate}, Nb) = (0.005, 5000).
Coordinates are given in arcseconds relative to $\alpha$\,=\,64.0382167, $\delta$\,=\,-24.0675012;
$e$ and $\theta$ are the ellipticity and position angle of the mass distribution.
Each parameter is given as the median ($\pm$\,1$\sigma$ confidence level) with the best fit value in parentheses.
For an L* galaxy, we have $\sigma$\,=\,208$\pm$8 km\,s$^{-1}$ and $r_{\rm s}$\,=\,15.0$\pm$2.3"\,{\rm{kpc}}.}
\label{bestmodel}
\end{table*}